# Planetesimals to Brown Dwarfs: What is a Planet?

Gibor Basri
*Astronomy Dept. MC3411, Univ. of California, Berkeley, CA 94720-3411*

Michael E. Brown
*Div. of Geological and Planetary Science, Calif. Inst. Of Technology M/C 150-21, Pasadena, CA 91125*

- **Abstract**    The past 15 years have brought about a revolution in our understanding of our Solar System and other planetary systems. During this time, discoveries include the first Kuiper Belt Objects, the first brown dwarfs, and the first extra-solar planets. Although discoveries continue apace, they have called into question our previous perspectives on planets, both here and elsewhere. The result has been a debate about the meaning of the word "planet" itself. It became clear that scientists do not have a widely accepted or clear definition of what a planet is, and both scientists and the public are confused (and sometimes annoyed) by its use in various contexts. Because "planet" is a very widely used term, it seems worth the attempt to resolve this problem. In this essay, we try to cover all the issues that have come to the fore, and bring clarity (if not resolution) to the debate.

## Introduction

Since prehistoric times, people looked into the night sky and picked out the planets. It was easy: they were the stars that moved (initially the Greeks considered the Sun and Moon to be planets too). Even when Galileo found that the planets are other worlds (and that the Earth is one too), there was no real controversy about what planets were. When Herschel spotted Uranus, he had no trouble claiming its planetary status (once its orbit was established). The real problems began at the start of the 19$^{th}$ Century, with the discovery of Ceres (right where the Titus-Bode Law said it should be). Although thought a planet for a few years, it was soon reassigned to "asteroid" on the basis of its size and mostly because of the company it kept (too many similar bodies in similar orbits were discovered). Neptune was never in doubt, and Pluto started off with only an inkling of trouble. As its estimated size shrunk (from like Mars to between Mercury and Ceres), people mostly shrugged their shoulders. When, however, other Kuiper Belt objects were discovered during the 1990's (in increasing numbers, including many "Plutinos" in the same resonant orbit and objects increasingly close in size to Pluto), the arguments against Ceres as a planet resurfaced against Pluto.

This last decade also saw the first discoveries of planets outside our Solar System ("exoplanets"). The first exoplanets discovered are actually terrestrial, but were found in orbit around a neutron star (Wolszczan & Frail 1992). These clearly have a very different history from planets in our Solar System, since their current orbits would have been inside the supergiant star that preceded the central pulsar. The next large group of (gas giant) exoplanets (well over 100 at this point), were discovered by the radial velocity variability they induce in their host stars (Marcy & Butler 1998). These objects are sometimes substantially more massive than Jupiter, and quite often found in surprisingly (compared with our Solar System) close and/or elliptical orbits. Recently, a few close-in exoplanets have also been found by the transit method - a drop in light caused by the exoplanet passing between us and the host star (eg. Burrows et al. 2004). Distant exoplanets have also been found by gravitational microlensing (eg. Bond et al. 2004).

Brown dwarfs ("failed" stars) were first discovered at nearly the same time, and then brown dwarfs with increasingly small masses began to be discovered (Basri 2000) . Inevitably, the puzzle of how to distinguish between exoplanets and brown dwarfs near some transitional boundary appeared. This question sharpened as very low-mass objects were discovered not orbiting anything else (eg. Lucas & Roche 2000; Martin et al 2001) . Systems were also found with exoplanets and brown dwarfs co-existing in "planetary" orbits (Marcy et al. 2001) . Astronomers and planetary scientists were forced to admit that they don't really have a clear-cut definition of "planet". Both these debates and the "Pluto controversy" came as a surprise to the public, who didn't realize there was any problem defining "planet" (after all, we unquestionably live on one!). We are just coming to the point where substantial numbers of terrestrial exoplanets can be found. The Kepler Mission (NASA) will utilize the transit method to accomplish this, beginning as early as 2008 (Borucki et al. 1997) . Presumably their identity will not be controversial (as they will most closely resemble our home world), though we should not underestimate the ability of Nature to hand us surprises.

It is valid to ask why we scientists should even care what the definition of "planet" is. Don't we just study whatever is out there? One answer is that we use the word quite a bit (as in the title of this publication), so it seems sensible to know what we mean by it. Another reason is that it is a "loaded" word. The public is very interested in planets, and learns about them as young schoolchildren. Certain kinds of discoveries (especially of new planets) bring benefits (or at least attention) beyond that for discoveries of similar objects that are not called "planets". . The latest example of this importance is the current (as of this writing) controversy around the discovery of the "10$^{th}$ planet" (2003 UB313) by MB and collaborators. There are sometimes professional arguments over whether the discovery merits such attention, based on the definition of "planet". We care in part because society cares.

*"When I use a word", Humpty-Dumpty said in a rather scornful tone, "it means just what I choose it to mean, neither more nor less".*
*"The question is", said Alice, "whether you can make words mean so many different things".*
*"The question is", said Humpty-Dumpty, "which is to be master, that's all."*

*Lewis Carroll in <u>Alice Through the Looking Glass</u> (1872)*

A good definition has several desirable characteristics. It should be succinct and easily understood by the public, yet precise enough to be acceptable to scientists. An ideal definition would not depend on specific knowledge or examples that will change as we learn more, though the definition of planet has already changed multiple times over the centuries of usage. Any definition should differentiate planets from other objects they might be confused with, preferably based on observables (to allow determination of whether a given object qualifies; it is even better if these observables are quantitative). One might hope that a definition not be time or history dependent (once a planet, always a planet), although there are proposed definitions for which that is not the case. There is another characteristic of a good definition that, although much less objective, also seems to be a major factor. The definition should have "cultural support", by which we mean that it cannot violate any strong preconceptions held by a large number of people. The word "planet" has been around for a long time, and scientists need to take care not to try to change too much the meaning of word that everyone already knows and uses.

Nature, of course, does not worry about definitions. Objects are made in various contexts, with a continuum of masses, and using a number of formation mechanisms. It is humans who wish to apply labels, and to distinguish between "different" objects. We must accept the fact that, in placing boundaries on a continuum, there might be transitional objects which are quite similar to each other in many respects, yet which are defined to lie on opposite sides of our imposed (and hopefully not too arbitrary) lines. A scientific definition would find natural physical conditions for such boundaries. A successful definition can only be arrived at through a consensus process. Our purpose here is to lay out many of the arguments and considerations which go into it. Whether some "official" definition is ever codified (or some consensus definition comes into wide use) depends on both scientists and the lay public having reached some common understanding.

One major step is to decide the arenas in which "difference" is to be defined. There are four basic arenas within which the planet definition debate is being played out. These are "Characteristics", "Circumstances", "Cosmogony", and "Culture". A good deal of the debate derives from the various weightings that different proposed definitions give to each of these arenas, and the rest of the debate arises from disagreements as to where lines should be drawn within them. In this article, we will examine each of the arenas, and discuss the issues and controversies they contain. This is not a traditional research review – it is more of an "op-ed" piece. We will assume that the reader has the background knowledge in planetary science to appreciate most of what we are discussing. It would take far too much space to provide all that background, so there is only a sparse bibliography, which is largely astronomical to bring current developments to the predominantly planetary audience. At the end, we provide a number of possible definitions for "planet".

# Characteristics: Physical Properties of the Objects

In astrophysics, the definition of objects is usually based solely on their physical characteristics, or observational characteristics if we don't yet understand the physical objects. A modern "star" is an object whose luminosity is derived solely from hydrogen fusion for some period of time. One need not specify how it formed, whether it is by itself or in a binary system or cluster, or if it dies as a white dwarf or in a supernova explosion. Prior to the 20$^{th}$ century, it was a "sun-like" object, and prior to that a fixed point of light on the celestial sphere.

For stars, the fundamental variable that determines the nature and fate of the object is mass. In the case of planets, the Earth is now unquestionably an example, but we argue about how to define the upper and lower mass limits (or whether to use some other scheme to set limits). We do know that if you keep reducing the mass, you eventually end up with "rocks" or "snowballs", and that if you keep increasing the mass you end up with "brown dwarfs" and then "stars". The question in this arena is whether there are observable, rational, informative, and/or compelling places to set the limits.

From a physical point of view, the limiting objects (which everyone agrees are NOT planets) differ in two different ways from planets near those limits. At the lower end, the size and shape of rocks are quite different from those of small planets (while the composition and luminous mechanisms can be similar). At the upper end, the amount and mechanism of the luminosity of stars is somewhat different from those of large planets (while the composition and size may be similar). It therefore seems sensible to concentrate on the physical consequences of pressure support versus gravity at the lower end, and the luminosity mechanisms at the upper end (while composition should be left out in both cases). Of course, the luminosity of stars is also a result of the competition between pressure support and gravity, so we are being consistent in that sense. There are other physical transitions which depend on mass that we can also consider.

## Effects of Increasing Mass

Let us now briefly review the competition between pressure support and gravity, from rocks to stars. Rocks are essentially supported by bound electron degeneracy (in atoms and molecules), while gravity plays no role in the size or shape of the object. This accounts for the bulk of macroscopic objects (like ourselves); without it everything would collapse to nuclear densities (with sizes more than 10 orders of magnitude smaller, as in a neutron star).

As one increases the mass of an object, several important transitions are reached. Perhaps the first is the mass at which gravity is able to hold together a "rubble pile". It seems that the comet (Shoemaker-Levy 9) that impacted Jupiter in 1994 and the comet Tempel 1 (determined by the Deep Impact mission) are such objects, along with some small asteroids (based on their densities). This limit is clearly too low (as nobody considers such objects planets). Next, the force of gravity can exceed the material strength of the body, and force the object to take the gravitational equipotential surface for its shape. That shape, of course, is a sphere (in the absence of rotation). One can estimate the size at which this transition occurs by equating the differential pressure of a deviation from sphericity with the yield or compressive strength of the material. For materials with the yield strength and density of stony meteorites the critical diameter is around 800 km.

This estimate is in accord with observations in the Solar System (Fig. 1). Ceres, with a average diameter of 941 km, is a gravitationally relaxed object (Thomas et al. 2005); although there is evidence it may have substantial water content in its mantle. The next largest asteroids, Pallas and Vesta, with approximate diameters of 608 and 538 km respectively, appear moderately elongated in HST images. Outer Solar System bodies made of an ice-rock mixture (more ice than rock) have lower yield strength and should become round at smaller sizes. Miranda is almost spherical at 472 km, but retains large topological features from an apparently violent youth; Mimas has a diameter of nearly 400 km but retains the huge Herschel crater. The largest known substantially non-spherical icy body is Saturn's satellite Hyperion, with dimensions of 180 x 140 x 112 km.

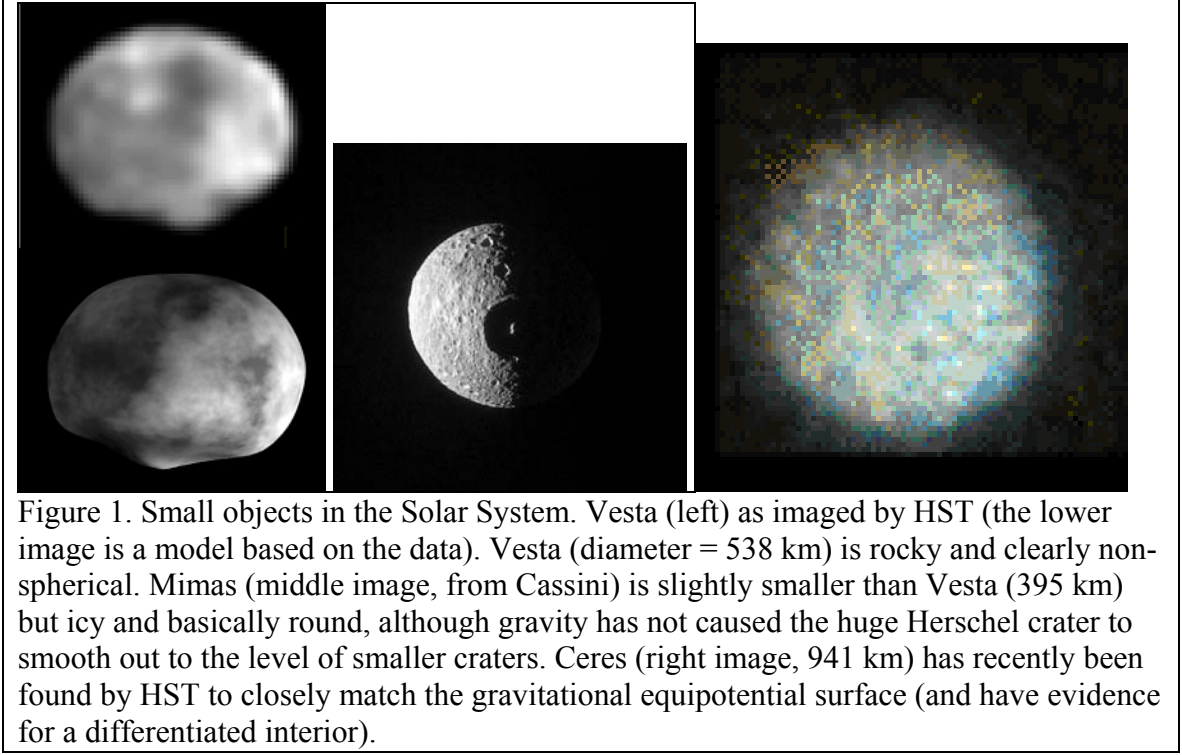

Figure 1. Small objects in the Solar System. Vesta (left) as imaged by HST (the lower image is a model based on the data). Vesta (diameter = 538 km) is rocky and clearly non-spherical. Mimas (middle image, from Cassini) is slightly smaller than Vesta (395 km) but icy and basically round, although gravity has not caused the huge Herschel crater to smooth out to the level of smaller craters. Ceres (right image, 941 km) has recently been found by HST to closely match the gravitational equipotential surface (and have evidence for a differentiated interior).

While becoming spherical is perhaps the most obvious outward sign of increasing mass, the interior of the body begins to undergo many interesting geophysical transitions as the mass increases. An important transition from an object which is geophysically dead to one that is active is the transition to bodies large enough to have convection in their interiors. The threshold for convection scales as the Rayleigh number in the interior which scales roughly as $D^5$ times the thermal contrast. This itself is likely to be an increasing function of D and a decreasing function of age –(for bodies of similar composition, where D is the diameter of the body). With such a strong dependence on planetary radius, solid state convection quickly becomes important somewhere in the size approximately between the Moon (D=3500 km) and Mars (D=6500 km). At a similar mass – a diameter of around 3000 km for a terrestrial body – the average gravitational energy per atom exceeds 1~eV, which is the typical energy of chemical reactions. A body this size has sufficient gravitational energy to substantially modify the initial chemical composition of its initial starting materials.

One final threshold that occurs near this size range is that when the central pressure of the object begins to be large enough that the materials have significantly higher densities than their non-compressed forms. The central pressure at which such compression occurs is approximately equal to the bulk modulus of the constituent material, which is approximately 1 Mbar for rocky materials and 10 kbar for icy materials. If we assume hydrostatic equilibrium and incompressible materials, the central pressure in a body is $P=(1.4 \text{ kbars})(\rho/1 \text{ g/cc})^2 (R/1000 \text{ km})^2$. Pressures high enough for significant compression are reached for a size of approximately 6000 km for rocky materials and only 1000 km for ice-rock mixtures. The true sizes needed for compression will be somewhat smaller, as the compression itself violates our incompressible assumption. The estimates of the body size for which these important transitions are reached are extremely rough and can be greatly influenced by composition, temperature, state of differentiation, and other details about the body. Nonetheless each of these transitions is conceptually well-defined and could equally well form the potential basis for drawing boundaries. After passing this mass range, there is not another significant qualitative change in the relation between pressure and gravity until masses greater than Jupiter. There was thought to be a significant "characteristics" boundary at about 10-15 earth masses, in the sense we didn't think we would find "terrestrial" planets with larger masses. This is because the supply of heavy elements in the protoplanetary disk is only something like one percent that of hydrogen and helium (which we will henceforth refer to as "gas"). It is hard to gather a more massive object without also gathering a lot of gas. The planets Uranus and Neptune are transitional objects in this sense. Very recently, however, the discovery of an exoplanet (Sato et al. 2005, Fig. 2) with nearly the mass of Saturn but only two-thirds its size has thrown this theoretical expectation into doubt (it may have a rock/ice core of 60 Earth masses or more).

For massive gas giants, a significant change in the source of core pressure support occurs. This change has been slowly brewing as one approaches higher masses, since thermal support begins to play an increasing role. The gas giants have very hot cores compared with terrestrial planets because the pressures are so large at their center, and this energy is able to free electrons even under such extreme pressures. Indeed, electrons were already becoming freer in the metallic hydrogen interior of a planet like Jupiter. At around two Jupiter masses, free electron degeneracy pressure begins to become comparable to "normal" pressure terms.

Until that point, adding mass to the planet made its radius grow. Free electron degeneracy is different in two fundamental respects. Firstly, it does not require thermal input to make the electrons move faster (thereby providing more pressure). They now do so because of the Fermi exclusion principle: there are enough electrons in a small volume that some can be forced to occupy very high energy levels simply because all lower ones are filled. This leads to the second difference, which is that producing increased pressure requires increasing density (not more heat). Thus, adding mass to objects supported in this way causes their radius to <u>decrease</u> (in order that their density increase).

The mass boundary where objects begin to become smaller and denser is at about twice the mass of Jupiter (Fig. 2). Such an object is thus the largest planet one can have, although the peak in size extends to about 5 Jupiter masses. A planet with ten Jupiter masses will actually be smaller than Jupiter, as are brown dwarfs. One caveat here is that we must wait for the outer layers to cool off before this is true, since they are not supported by free electron degeneracy. This is why the "hot Jupiters" shown in Fig. 2 are actually larger than the peak size of cold planets. We do not recover the usual case of size increasing with mass again until we reach true stars, whose nuclear fusion can get the core hot enough to lift the free electron degeneracy. The onset of degeneracy (the largest size reached by non-fusing objects) provides one natural place to draw a boundary.

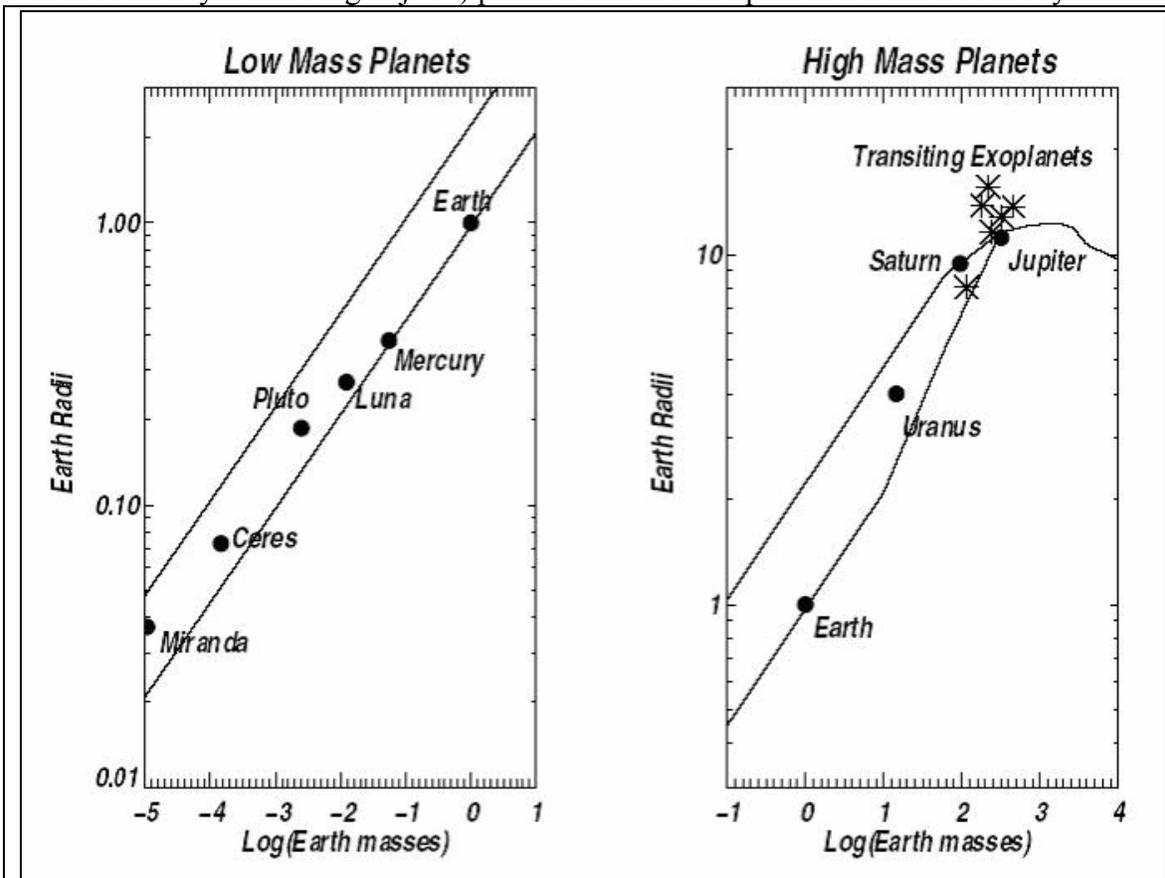

Figure 2. The mass-radius relations for planets. This is essentially just a statement of bulk densities. For low mass planets, the lower line represents a bulk density of 6 gm/cc, while the upper line is for 0.5 gm/cc. The planets lie between these lines depending on their composition: rocky planets with high concentrations of heavy elements have high densities, and icy or gaseous planets have low densities. The relation changes at Jovian masses because degeneracy pressure plays an increasing role (the density of the planets begins to increase rapidly). Furthermore, it is difficult to construct very massive planets with primarily heavy elements (the behavior shown for the high density case is arbitrary between Uranus and Jupiter). The mass-radius relations are only known for extrasolar planets (asterisks) that transit their stars and have radial velocity data (giving the radius and mass respectively as a fraction of the stellar values).

## Sources of Luminosity

We do not discuss luminosity sources for terrestrial planets, as nobody has proposed using them to define limits at low masses. For gas giant planets, the very process of cooling (radiating luminosity) allows the planet to contract, releasing gravitational energy. This is the dominant source of luminosity for jovian planets (and indeed for brown dwarfs, and stars in their pre-nuclear burning phases). The luminosity-size history of the object can be simply derived by knowing how easily it radiates contraction energy away. The contraction continues until another pressure term (free electron degeneracy for very large planets) halts it. There is no nuclear fusion at all.

The minimum mass required to cause substantial fusion is about 13 Jupiter masses. What is fused at that mass is "heavy hydrogen", or deuterium. All bodies form with a supply of deuterium (produced in the Big Bang) at about $10^{-5}$ the abundance of light hydrogen. All stars and brown dwarfs burn their deuterium during the early part of their life; more massive objects burn it faster. For objects near the fusion boundary, the deuterium-burning phase lasts more than 50 million years (much longer than the total fusion lifetime of very high-mass stars). At the boundary mass, the deuterium burning does not fully halt the contraction or contribute most of the luminosity at any time, but for just a little higher mass (like 15 Jupiter masses) it basically does. The fusion limit constitutes a well-defined qualitative and measurable (in principle) difference between objects that are otherwise quite similar.

## Circumstances: The Context of Planets

We now discuss the circumstances under which an object should earn the "planet" label. It is well to remember that the question of whether orbits themselves are a requirement is one of tradition (as we discuss in our "Culture" section). The requirement has been in place for the last 4 centuries or so (but not before then!). If Mercury were orbiting Jupiter rather than the Sun, it would clearly never have been considered a planet (as is the case for Ganymede), while if our Moon were in Mercury's place, it would be (barely) observable with the naked eye and would have been considered a planet even before the Copernican Revolution. Clearly, the common use of "planet" includes orbital circumstances.

Other circumstances to be considered include whether the planet is "unique", or clearly a member of a population (defined dynamically). The question of whether the orbit fits our Solar System intuitions also keeps coming up. We will also see below that circumstances can change. This immediately raises the question of whether it is the current circumstances that matter, or the full history of circumstances.

### Orbital Dynamics of Planetary Systems and Protoplanetary Disks

We first discuss the situation when planets are forming. Initially they will do so inside a massive gas disk (the ratio of gas to dust is generally 100:1 in the interstellar medium).

When particles are very small (up to the size of chondrules or so) they are subject to Epstein drag with the gas. As they grow bigger, they are still subject to gas drag, which arises because the gas has a slight radial pressure, causing it to orbit slightly more slowly at a given radius. This causes bodies up to meters in size to drift rather rapidly inward, especially in the inner planetary system, posing one of the several problems in understanding how interstellar dust can reach planetesimal size (kilometers).

There are several basic types of orbital migration that result from the interaction between a forming planet and the gas disk (cf. Nelson & Papaloizou 2004) . Two are due to the production of waves and torques in the gas disk by the planet (they are usually called Type I and Type III). They tend to operate on bodies with masses between lunar and a few Earth masses, and drive the bodies rapidly inward (on timescales short compared with the disk lifetime). Another is due to the opening of a tidal gap in the disk by a massive planet (Type II). In this case the planet gets locked into the disk, and shares its motion thereafter (inward drift if in the accretion part of the disk). The timescale here is longer, but still of concern. All mechanisms tend to result in planets falling into the star (or getting parked very close by, perhaps). On the positive side, this could explain the presence of massive exoplanets close to their stars. On the negative side, it is actually hard to retain planetary cores long enough to induce runaway gas accretion and build giants. Of course, the outer regions of the disk are actually spreading, so planets locked out there will migrate outwards. The disk viscosity that drives all this is rather poorly understood, so this topic remains one of active research.

A related topic is the sweeping up of planetesimals (or smaller planets) as a giant migrates inward, and another is the capturing of such bodies in outer orbital resonances as they try to catch up to the giant. Both of these depend (in opposite ways) on whether the smaller bodies migrate faster than the large body. Much theoretical effort is going into understanding these migrations better, and with more sophistication.

Disks of planetesimals can also drive orbital modification. Saturn, Uranus, and Neptune migrated substantial distances outward as they scattered planetesimals predominantly inward. Jupiter migrated slightly inward as it removed many of these planetesimals from the Solar System. Dynamical friction can also play an important role in the types of orbits that are produced and in the later mergers of planets and planetary embryos to produce the "final" population. Orbits also can be changed through direct gravitational scattering. Particularly when at least one of the planets is massive, other planets can have their orbits substantially altered, or even be ejected from the system. If there are several massive planets, quite a bit of system evolution can occur. It is clear that the orbits of the planets can change dramatically during these early phases.

The next, and perhaps most important, transition occurs between objects which are or are not massive enough to clear residual planetesimals through accretion or scattering. Empirically we can see that Ceres, for example, is not massive enough to clear the asteroid belt of its many remaining planetesimals, while the other terrestrial planets have certainly been successful in clearing their regions of influence. As a cultural matter, collecting a few small bodies into orbital resonances has not disqualified an object from being a planet (eg. the Trojans for Jupiter or the Plutinos for Neptune).

The most stringent criterion for clearing planetesimals is that the surface escape velocity – which is the maximum velocity that one body can impart on another through gravitational interaction – be greater than the local escape velocity from the central star. A body of such mass will be able to scatter other small bodies beyond the gravitational influence of the star. This criterion is met when the ratio of the planet to central star mass is greater than the ratio of the planet's distance from the star to the radius of the star. In the Solar System this threshold is only crossed for Jupiter, which is thus the only planet capable of easily removing bodies from the Solar System.

For bodies less massive than Jupiter, clearing proceeds through a combination of accretion and the eventual influence of larger bodies. A strict criterion for the size at which this clearing happens is difficult to decide upon. One could argue that the terrestrial planets would not have been as effective in clearing planetesimals without the influence of Jupiter. In the outer Solar System, for example, planetesimals have mostly been cleared inside of Neptune, where they can eventually be influenced by Jupiter, but not outside of Neptune, where they have been pushed just to the limit of Neptune's influence. Empirically, however, it is easy to determine the largest bodies in the Solar System that have not cleared their region: Ceres is still surrounded by the asteroid belt, and Pluto and 2003 UB313 are still surrounded by the Kuiper belt.

To roughly estimate what mass Ceres or 2003 UB313would have to have to clear their orbital neighborhoods, one can examine the spacing of the "true planets" in terms of their Hill radii (the "Hill sphere" is defined roughly as the size of the region around a planet where its gravity dominates that of the central star). The terrestrial planets that have cleared the planetesimals between them are separated by approximately 30 Hill radii each, while the giant planets are separated by about 10 Hill radii. An object in the asteroid belt that cleared the region between that object and Mars would need about 1 Earth mass. In the Kuiper belt, coincidentally, a similar mass is required. While these estimates are rough, they suggest that the largest belt objects, 2003 UB313and Ceres, are not massive enough to clear out their populations by about 2-3 orders of magnitude (more detailed calculations agree; Stern & Levinson 2002).

The "clearing" criterion divides the Solar System into two distinct classes. The eight largest bodies have all cleared residual planetesimals from their vicinities, while the others have not. This criterion is the basis for one possible dividing line between objects that are planets and objects that are not. A planet could be by definition an object that has cleared residual planetesimals, and thus cannot be part of a population of smaller bodies.

As with all definitions, this one has advantages and disadvantages. The main advantage is that it correctly divides the Solar System into its dynamically important constituents. Dynamically, there is no argument whether Ceres and 2003 UB313naturally belong to the asteroid and Kuiper belt populations rather than to the collection of larger objects. One disadvantage is that there is no simple a priori mass threshold above which an object would be classified a "planet". An interesting situation would arise if, for example, a Mars-sized object were found in the inner Oort cloud. By the dynamical definition this object would not be considered a planet, yet most people would instinctively feel that a body that size should indeed be called a planet.

For extrasolar planetary systems there are likely to be a wide variety of unforeseen dynamical circumstances, and those that have been seen have mostly been surprising. Many objects are found in eccentric orbits, some in mean motion resonances, some with absurdly short orbital periods, some with uncomfortably long ones (given our Solar System preconceptions about how they form). Because of our current mass-detection limits, all known exoplanets are well above the mass needed to clear planetesimals from their neighborhood. Interestingly, few of the objects currently known are massive enough to directly eject planetesimals, because of detection biases favoring short-period planets that are deep in the central stars' gravitational wells. Of particular interest will be new systems with terrestrial planets and no giant planets.

Are there any dynamical circumstances that should preclude an object from being considered a planet in the circumstantial arena? As we did for increasing mass, we search for natural dynamical transition points, and find only three. The first is between bodies that are on stable orbits and those on unstable orbits. Even this distinction is ambiguous. The Solar System itself is unstable on sufficiently long time scales, and it is likely that planetary-sized bodies have been ejected from it in the past. All planetary systems after their formation phase will be, by definition, at least marginally stable on the timescale of their current lifetime, so there is no real need to distinguish between stable and unstable systems. The second transition point, which we have concentrated on, is whether the object can clear its neighborhood of smaller bodies. We have seen that this does not depend only on mass, however, but also on circumstances (past and present).

The final and most obvious dynamical distinction is between bodies that are bound to a massive central object (or objects) and those that are not. From the point of view of orbital dynamics, at least, it makes no sense to talk of planets unless they are – or were – bound. One question is whether the central object has to be a star, or whether a brown dwarf (or neutron star) will do; most people seem willing to include all objects capable of fusion. But if isolated massive non-fusing objects are not planets because they don't orbit stars, what are the smaller objects that could be orbiting them? Moons? Planets? Dynamical definitions also get into trouble when discussing ejected planets. Is an object that was once orbiting a star and had the attributes of a planet, but is then ejected, no longer a planet? It seems that the history of an object could play a role in whether it is a planet, and that brings us to its earliest history, or cosmogony.

## Cosmogony: The Formation of Planets

This arena holds great importance for many scientists (though less so for the public). It is perhaps the source of greatest disagreement among astronomers. There is a large segment who hold that "planets" are by definition objects which "form in a disk around a central star [perhaps] from the accumulation of planetesimals," and that is the most important defining property they have. The pure form of this position insists that if the planet grows past the fusion limit, it is still a planet. Presumably this position also implies, if it turns out that Jupiter has no real core and formed suddenly during a rapid gravitational instability in the protoplanetary nebula, that we have been mistaking a "sub-brown dwarf" for a "planet" all this time. Such a claim, of course, would run afoul of the "culture" arena. This difficulty is avoided by dropping the insistence on formation through the specific mechanism of the accumulation of planetesimals.

The other element insisted upon is that the nucleation of the planet must take place in a disk around a star (many accept a brown dwarf as a central object as well). Objects (of any mass) that form in the centers of isolated disks (even if they ingest many planetesimals) are not "planets" in this view. This "cosmogonist" group of scientists is in contrast to another large group who think "characteristics" should receive greatest weight in the definition, and a third group who also give a lot of weight to "circumstances". The traditional position on planets (before 1990) is an amalgam of all three arenas, heavily informed by our known Solar System at that time (culture).

The presence of disks in the formation of planetary (and stellar) systems is a given in Nature (cf. Mannings, Boss, Russell 2000). There is essentially no chance that a collapsing cloud can contain so little angular momentum that this won't happen. The question is how massive and large the disk will be relative to the central object, and whether it will be conducive to the subsequent formation of planetesimals and planets within it. Evidence is quickly building that grain growth in disks is early and robust, although the transition from large grains to planetesimals is much less understood. Once planetesimals are present in large numbers, the path is clear to larger bodies through pairwise accretion in what is called "oligarchic growth" (larger bodies grow faster than smaller bodies – a runaway process). There are further mergers between protoplanets and/or planets, and some may be ejected from the system altogether.

In a circumstellar disk, objects above about 1 km in size can first begin to cause enough gravitational perturbation to affect the dynamics within the disk. At this size we first call the bodies "planetesimals," and the combined effects of gravitational focusing and velocity evolution eventually causes runaway growth, to the sizes of "protoplanets". In the simplest analytic planet formation scenarios, these are objects that have reached the "isolation mass". That is, each protoplanet has swept out all of the material within its Hill annulus and is accretionally isolated from the other protoplanets. For the minimum mass solar nebula at 1 AU, this isolation mass is approximately equal to the mass of Mercury. In the Kuiper belt region, the Hill annulus is significantly larger and the isolation mass is approximately 7 Earth masses. Any objects more massive than protoplanets must have undergone mergers, and are not pristine protoplanets any more.

There is also an agreed upon "standard model" for gas giant planet formation. This requires the growth of non-gaseous planets up to 10-15 Earth-masses while the protoplanetary disk still has its gaseous component. Astronomers have found that gas disks typically last around 5 million years, with the oldest lasting perhaps twice that long. This sets the timescale over which the standard model must operate. When the planetary core grows that massive, there is a rapid runaway gas accretion phase. As the object grows it opens a tidal gap in the disk, but it is now fairly clear that accretion can continue even after the gap is opened (accretion streams spill into the gap). The size to which the gas giant can grow is controlled by many factors (which vary among systems), and is not really known.

The main competition to the standard model comes from "rapid collapse" scenarios, which posit that disks that are cool and massive enough can form gas giants directly through gravitational instabilities (Boss 1997). This resembles the means by which close stellar binaries form. They also form "in disks"; typically there will be a circumstellar disk around each object, and often a circumbinary disk around both of them (depending on their separation and the total angular momentum present). As one shrinks the size of the secondary compared to the primary, the circumbinary disk grows relatively larger and the disk around the secondary shrinks. The system's appearance slides smoothly over to what can be described instead as a small object (surrounded by its own disk) embedded in a tidal gap in the circumstellar disk of the primary. This picture is the same as in the rapid collapse scenario, and its resemblance to binary star formation is one of the main factors invoked by those in the cosmogony camp against including it for planets.

Of course, a small object surrounded by a small disk embedded in a gap in a much larger disk is also a description that applies to a gas giant forming in the standard scenario. There is clear evidence in the nature of the Galilean moon system that Jupiter had such a disk. The time needed to form a planet is much less in the collapse scenario, which also results in the lack of a denser planetary core. The rapidity is one of its most attractive features; it avoids the problems due to orbital migration and the dissipation of the circumstellar gas disk.

The collapse scenario is less well-accepted for a number of reasons. The main empirical reason is that we know that 3 of our 4 Solar System giant planets have cores, and Jupiter likely does or did too, giving no cause to look for an alternate mechanism. The main theoretical reason is that the conditions required in the disk to cause the gravitational instability are rather extreme, and it has not been shown that Nature will actually produce them. Well-mixed disks are difficult to cool to the point of instability (Bodenheimer 2004). One way around this is if the dust settles out, making a cold dense layer at the disk midplane. In principle, some of the gas layer could be depleted (by photoevaporation for systems in clusters with high-mass stars, for example, or by the stellar wind), also producing some metallicity enhancement relative to the central star. Mitigating against this are turbulent forces in the gas which might keep the dust mixed in, and this problem is not well-understood enough to make definitive conclusions currently.

There is also a scenario under discussion to bridge the grain-to-planetesimal gap by gravitational instability (Youdin & Shu 2002). In a sufficiently settled dust disk, it may be possible to get gravitational instabilities that can produce small (10-100 km) bodies, which will be rock/ice rather than gas. This would be somewhat ironic, since one might end up insisting that gas giant planets must be produced by gathering planetesimals and not by gravitational instability, but the planetesimals themselves could be formed by direct collapse. Until there is more convincing evidence for gravitational collapse at the right mass scales, however, it is possible to dismiss this part of the discussion as purely theoretical.

There does seem to be one case where such evidence for direct gravitational collapse of planetary mass objects may already exist, and that is the case of what have been called "free-floating planets" (eg. Lucas & Roche 2000; Martin et al 2001). Isolated objects are found at increasingly low masses, and some are apparently below the fusion limit. There is no theoretical reason why the star-formation process (rapid collapse from an interstellar cloud) should suddenly fail at the fusion limit. This is one of the arguments given for insisting on planetesimals to define planets: do we want to suddenly change the name of objects forming by direct collapse just because one (by now minor) source of luminosity (fusion) has given out? Another type of evidence for the collapse scenario is discussed below – giant exoplanet candidates imaged far from their stellar hosts.

## Culture: The Common Understanding of "Planet"

Until the 1990s, there was not much controversy about the definition of "planet". It had been revised at the end of the 17$^{th}$ Century to its current meaning (many dictionaries still retain the astrological understanding of the word too). We start our section on culture by looking at a few dictionary definitions, which together capture much of the general public impression.

The Oxford English Dictionary offers:
> *Mod. Astron.* The name given to each of the heavenly bodies that revolves in approximately circular orbits round the sun (***primary planets***), and to those that revolve round these (***secondary planets*** or ***satellites***).
> The primary planets comprise the ***major planets***, of which nine are known, viz., in order of distance from the sun, Mercury, Venus, the Earth, Mars, Jupiter, Saturn, Uranus, Neptune, and Pluto, and the ***minor planets*** or ***asteroids***, the orbits of most of which lie between those of Mars and Jupiter. [Emphases are in the original]

Dictionary.com offers:
> A non-luminous celestial body larger than an asteroid or comet, illuminated by light from a star, such as the sun, around which it revolves. In the solar system there are nine known planets: Mercury, Venus, Earth, Mars, Jupiter, Saturn, Uranus, Neptune, and Pluto.

Merriam-Webster (online) offers:
> **b** (1) **:** any of the large bodies that revolve around the sun in the solar system (2) **:** a similar body associated with another star

It is interesting to note that OED seems to include asteroids, and even moons, in the general definition, and also includes "approximately circular orbits" (a nod to cosmogony, unwitting or not). Apparently neither the Kuiper Belt nor exoplanets have worked their way into this exalted source yet. The other sources want planets to be "large" (bigger than asteroids or comets). There is an instance of a requirement on luminosity sources. Pluto is definitely included in the first two sources, but the third has a less Solar System-centric point of view. All require that planets orbit stars, and this seems to be the strongest agreed-upon characteristics of planets wherever you look.

Culturally, there are a few additional important planetary characteristics that appear to have near universal support. Not surprisingly, all are difficult to quantify, but most people have no difficulty understanding them. At one time, many people might have suggested that planets need to be in orbits similar to those of the Solar System. The fact that few worry about this aspect of exoplanets anymore is an interesting indicator of how fast cultural perceptions can (and should) change in the face of new information. One criterion that most people appear to hold is that planetary status should be special in some way. In the Solar System the number of planets is sufficiently small that many people know them all. Any planetary definition that radically increases the number of planets in the Solar System will destroy this important cultural aspect of planets. In extrasolar planetary systems we are unlikely to reach the point anytime soon that we are in danger of violating this criterion. We are already comfortable talking about extrasolar debris disks and their parent populations of small bodies.

A book is about to appear which treats the cultural history of the word "planet" in far more detail than we can here. It is titled: <u>Is Pluto a Planet? A Historical Journey through the Solar System,</u> by David Weintraub (Princeton Press), and it contains a superb summary of all the historical meanderings and events that have led to the current situation. You might be surprised to learn of (or be reminded of) all the definitions and planets themselves that have been proposed and withdrawn over the centuries. The book also treats current discoveries, and this debate, in some detail.

The Pluto Controversy

Within the Solar System, the question of "what is a planet?" has recently resurfaced with the discovery of Kuiper Belt Objects (Jewitt & Luu 1993). When it became clear that Pluto was not an odd isolated object on a unique orbit, but simply one of the largest known of a substantial population, some scientists invoked a "circumstances" argument and suggested that it is now wrong to still call Pluto a planet. They suggest that it should be reclassified as the second largest known KBO. A similar reclassification happened with Ceres and the asteroids once it was realized that they, too, are members of a much larger population.

Traditionalists object that today Pluto has strong cultural status as a planet in a way that the asteroids never had in the early 1800s. The "characteristics" planetary definition, whereby all spherical primary objects orbiting stars should be considered planets, poses a conundrum for traditionalists. It retains Pluto as a planet, but also Ceres and perhaps tens to hundreds of KBOs. This has the cultural advantage of saving Pluto, but that is strongly offset by the cultural disadvantage of suddenly increasing the number of planets by as much as an order of magnitude. As noted above, sphericity is just one of the transitions that occur with increasing mass. The more geophysically important boundaries are at significantly higher mass, and would exclude all of the asteroids and known KBOs.

Pluto's situation clearly shows that there is no way to be scientifically consistent and also satisfy tradition in this case. The dynamical definition, while scientifically consistent, flies in the face of the apparent cultural desire to retain Pluto in the pantheon of planets. The spherical definition, while scientifically consistent and easily applied, does perhaps more damage to tradition than the simple reclassification of Pluto. No one has developed a logical definition of "planet" that retains the nine historical planets and adds no new ones. There are only two solutions: give up on culture or give up on scientific consistency. For scientists the inclination is to regard consistency as more important than culture (GB's position), but in the realm of planets where culture has a serious impact, it is worth considering giving up on consistency (MB's position). A final alternative would be to compose a consistent general scientific definition, but agree to apply a cultural definition within our Solar System.

## Isolated Objects Without Fusion

Beyond the Solar System, one of the reasons that the debate on "what is a planet?" arose is the discovery of isolated objects that are probably, but not certainly (due to current imprecision in determining their mass), incapable of fusion. The original discoveries were dubbed "free-floating planets" by those in the "characteristics" arena. Cosmogonists objected that such objects were probably not formed "like planets" (although this is not an empirical claim). Clearly, however, if planets must be in orbit around stars, those in the circumstances arena have a valid empirical objection.

Modern star formation theory muddies the waters by finding that objects can form as part of a "small-N" cluster, whose components interact gravitationally for a while, but where the lightest objects will be ejected from the group first. One might then view them as "ejected planets", although they were never in really stable orbits. We have come to see, however, that the concept of "stable orbits" is questionable even in the young Solar System.

## Objects in Unexpected Orbits

It has become apparent during recent announcements of "images of exoplanets" ( that circumstantial arguments are causing some scientists to reject the objects' planetary status. The objects found so far have at least several Jupiter masses (Neuhäuser et al. 2005), and are seen 50-100AU from their stars (or in one case, from a brown dwarf; Chauvin et al. 2005). Some astronomers reject these objects as planets because their distance from their host stars suggests that they may not have formed in the standard scenario (which is not thought to operate rapidly enough out there to beat the short lifetime of gas disks). They are being called "sub-brown dwarf binaries" and similar names.

This is an example of circumstance being invoked with a subtext of cosmogony. Presumably this objection would be mitigated if the objects were found to be on highly eccentric orbits, suggesting that they obtained their large distance by gravitational scattering with inner giant planets. It will be difficult to establish the orbital parameters anytime soon, however. Gaps and warps are being detected in debris disks at similar distances from their stars, which giant planets are invoked to cause. Discomfort with these distant giants, which would not have highly eccentric orbits, has not yet been expressed much (perhaps because the putative planets haven't actually been seen).

It is worth recalling the time of the first exoplanet discoveries (Marcy and Butler 1998). These have Jovian masses, but are in very close orbits to their stars. Gas giant planets simply weren't supposed to be found there (no formation scenario works in so close). As planets slightly further away were found (where tidal circularization no longer could operate), their orbits were also found to be mostly eccentric (similar to binary stars at similar separations). A few astronomers objected that whatever was being found, it was not planets, because massive planets don't have such orbits. There was initially the possibility that brown dwarfs were being found with low orbital inclinations (radial velocity only provides lower mass limits), but this was soon removed as more objects were found. Fairly quickly, however, we came to understand that eccentric giant planets could be found close in, and we had been biased by our previous reliance on only the Solar System. This "objection on the basis of expectation" is being applied once again to the distant massive objects. One wonders whether objections to planetary status would still arise if the distant objects had less than a Jupiter mass (despite the fact the standard model would still have a very hard time producing them).

## Defining Planets

We have now presented all of the reasons we know why the definition of "planet" has become a contentious and difficult-to-solve dilemma. It is clear that agreement cannot be reached on a definition before a consensus on the arenas of the definition is reached. Here the community seems divided into two broad camps. Those for whom tradition is less important, while logical consistency and empirical verification are most important, lean towards characteristics. They tend not to give the Solar System as great a weight compared with other planetary systems, and feel that recent discoveries call for a reopening of the whole question. The purists in this camp reject any consideration of circumstances, cosmogony, or culture – they prefer a purely mass-based definition. The primary problem for this approach comes from culture; there is not a critical mass of cultural support for that position.

Using the influence of gravity to draw the lower mass limit places uncomfortably (by tradition) small objects into the planetary domain. The number of known planets in our Solar System instantly rises to something like 12-15 (with the likelihood of many more to be found), and the extra ones are in the belts (eg. Quaoar; Brown & Trujillo 2004). Such a state of affairs is unacceptable to those who believe that the clear circumstantial differentiation between planets and objects in populations should be the basis for a division.

It is also unacceptable to those for whom tradition (culture) is most important. They are greatly influenced by the paradigm we have developed over the last few centuries. They want a definition that preserves both the current list of planets, and our theories about how they were formed. Objects that do not fit comfortably into these requirements should get other names, and definitions that seem to violate them should be rejected. This can apply to all three arenas: planets should not be too small or too massive compared to the traditional list, they should orbit stars (without too much company), and they should be formed from planetesimals in disks. It may be that a majority falls into this camp at the moment (but no proper polling has been done).

The difficulty with the traditional position is in the details. No particular weighting among the arenas has found general agreement. The Pluto problem poses a substantial obstacle. Either tradition or logical consistency must be abandoned. It also seems parochial to base a general definition of planets too closely on what we happen to see here; it is quite clear by now that other planetary systems may vary widely from the familiar. It is not clear when exoplanet discoveries should be allowed to dislodge long-held beliefs derived from the Solar System. Alternatively, one could argue that we will always know more about the Solar System than any other planetary system, so basing definitions on the best-studied examples might be wise.

Insistence on a particular cosmogony is also problematic. It is not empirically verifiable for most exoplanets; their histories remain largely out of reach. There may be an overlap between formation mechanisms operating over a certain mass range; there could also be an overlap between planets and brown dwarfs that form with a given mechanism. We do not really know enough about planet formation to be confident on these points. Purists in this arena, however, are willing to give up verifiability. They are also willing to call fusing objects "planets" if formed by nucleation in a disk, even if they have the same mass as brown dwarfs, and non-fusing objects "sub-brown dwarfs" if they formed by direct gravitational instability (whether in a disk or not). In some cases they would be content to say that an object's planetary status can't be confirmed or refuted empirically. This approach is perhaps more philosophical than scientific (though science used to be natural philosophy).

We now become more specific about proposed definitions. MB invokes doses of cosmogony and circumstance, with a strong nod to culture. GB has a preference for characteristics, with a sprinkle of culture.

*"That's a great deal to make one word mean", said Alice.*
*"When I make a word do a lot of work like that", said Humpty-Dumpty, "I always pay it extra."*

*Lewis Carroll in <u>Alice Through the Looking Glass</u> (1872)*

## Mike Brown's Discussion

Cosmogony is for me the most satisfactory method for defining the difference between planets and stars/brown dwarfs, even if it is the hardest method to actually use. In simple cases, like a single star and low-mass planet for example, it is trivial to distinguish between the star (which forms by collapse from an interstellar cloud, even if surrounded by a disk after a while) and the planet (which is built later in the star's circumstellar disk). Planet definitions involving cosmogony get into trouble as the mass of the secondary object increases. When is the secondary object a planet and when is the system a double star? We have discussed above the cosmogonies that are usually meant for planets, and how they differ from "star-like" formation. This scheme works in the Solar System, but is difficult to apply empirically for exoplanets.

It seems we are left with an impractical task, but Nature has apparently provided a simple and remarkably powerful observational solution to this difficulty of separating stars and planets by cosmogony. Radial velocity studies have found that objects in orbits within a few AU with masses no larger than Jupiter are fairly common, but the number decreases sharply as mass increases. Between roughly 5 and 60 Jupiter masses, the mass regime known as the "brown dwarf desert", there are few close stellar companions known (yet they would be easier to find). Above that mass, the realm of double stars is again well populated.

Nature has divided close stellar companions, by some mechanism, into low and high mass populations with a wide gap between the two. Such a clear dividing line invites a classification scheme. Jupiter and the rest of the Solar System planets fit firmly within the low mass population. It is natural to call this low mass population "planets" and the high mass population "stars." Note that there need not have been such a natural dividing line; the two populations could have merged smoothly (putting us back into the cosmogony conundrum). As we learn more about the populations more distant from the central star we may have to change our ideas about where and how to draw planetary dividing lines. The reason their status is uncertain is because we truly are uncertain of their provenance, and uncertainty about what to call them is perhaps the only honest response. We should feel no embarrassment about learning more and modifying what we previously meant when we said "planet."

To set the lower mass limit of planets, which for now only matters for the Solar System, I prefer the dynamical definition that the object has cleared planetesimals from its neighborhood, and thus is not part of a population. Any view of the Solar System that sees Pluto or 2003 UB313 as tiny lonely outposts on the outer edge, or Ceres as the sole mini-planet between Mars and Jupiter, misses one of the more important points about the architecture of the Solar System and the importance of its belts. The "clearing" definition appears to me to be the closest one can come to meshing the cultural and scientific views of what should be called planets. It is consistent with the arguments from the past reassignment of Ceres from planet to asteroid. The only casualty to the currently accepted system is that Pluto must go the way of Ceres and be likewise reassigned. It seems a minor price to pay for a definition that is close to satisfying culture while remaining scientifically rigorous.

Experience with the public after the discoveries of Quaoar and Sedna, however, has slowly taught me not to ignore the importance of culture. Those in favor of absolute scientific consistency regardless of cultural beliefs (which, until recently, included me) should also argue that Madagascar, an island sitting on an isolated block of continental crust in the Indian Ocean, should be called a separate continent. Though their arguments would be scientifically sound, they would not get very far, even with geologists. Likewise it has become apparent in the past few years that the extent of education and media makes it unlikely that the idea of Pluto as a true planet will ever be abandoned. One can argue with culture forever and, apparently, get nowhere. I thus finally concede: Pluto is a planet because we say it is and for no other reason. If need be, we can give Pluto an adjective and call it an "historical planet." All new Solar System objects bigger than Pluto join the planet club by default. 2003 UB313, a little larger than Pluto and spectrally similar, is a planet. 2005 FY9, a little smaller than Pluto but spectrally similar, is not.\ This one hundred percent cultural definition requires scientists to relinquish their desire to legislate a new and rigorous meaning to the Solar System sense of the word "planet" and accept the meanings that culture has been using all along. Planets are far too important to culture to expect that any newly legislated definitions will have much sway.

**Definition**: (a) *a planet is an object that is massive enough to clear planetesimals from its orbital neighborhood, and which is part of the empirically defined distinct group of low mass stellar companions with masses lower than about 5 Jupiter masses (b) in the Solar System, a planet is any of the nine historical planets plus any newly found objects bigger than the smallest of these*.

## Gibor Basri's Discussion

I have summarized much of my thinking in an article in Mercury magazine (Basri 2003). Most objects in astrophysics are defined on the basis of their physical characteristics, and now that planetary science is moving beyond the Solar System, that seems like the right thing to do here as well. I do not favor cosmogony as part of the definition for planets at all, partly because I don't think we know what we are talking about to a sufficient level of understanding. More importantly it is almost impossible to apply empirically outside our Solar System. I don't accept a situation in which a 4 Jupiter mass object seen 40 AU from a solar-type star might actually either be a planet or be something else (one could get that circumstance with several different histories). I don't think that the scientists who are trying to study such objects (and the many other odd exoplanets) will stand for being told that there is little hope of deciding whether they are studying planets or not.

Similarly, I don't much care for circumstantial definitions. I view them as just that, and not related to a fundamental class of astrophysical object. In particular, I think it is time to drop the great observational bias provided by our Solar System. Just as we should not think the current crop of detected exoplanets is representative of exoplanets in general, there is no reason to think that the Solar System is representative of planetary systems in general.

I do find it expedient to make one exception. That is because of culture – the fact that there is wide agreement that planets should orbit stars (which is a circumstance). My definition, therefore, is given in two parts. One is purely mass-based (I would call it the astrophysical part), and the other takes account of the one circumstance I can't see culture doing without. My mass-based limits are gravitational shaping at the lower end, and the ability for fusion at the upper end. I would say that these define the domain of "planetary mass objects", which I have shortened to "planemos". I call all objects above this mass range "fusors" (a discussion of "what is a star?" was not commissioned here). Then planets are just planemos orbiting fusors. This scheme includes Pluto and 2003 UB313 as planets, but also Ceres, Quaoar, Sedna, and other yet to be discovered and/or named KBOs that are sufficiently round. The recent work on Ceres (Thomas et al. 2005) only reinforces the idea that Ceres has all the properties of miniplanet as defined here.

To deal with the cultural desire that planets be special, I greatly favor the use of adjectives to distinguish between important classes of planets. I would call planets that are not dynamically dominant "miniplanets", thus diminishing the planetary status of Pluto without eliminating it (and forcing logical consistency with the other miniplanets). I would deal with the cosmogony argument (invoking some of Mike's reasoning) by calling planets whose cores are becoming degenerate (those more massive than about 2 Jupiter masses) "superplanets". This places all the objects whose provenance is easily questioned into a separate category, without fully robbing them of their planetary status. It does not reduce the current count of exoplanets.

Free-floating objects in the right mass range are all planemos, but not planets, and there is no need to determine their cosmogony to classify them. Some might indeed be ejected planets, but one would not call them that unless there were empirical reasons for doing so. There are no (insuperable) empirical obstacles to determining the status of objects in this scheme, and much of the cultural heat could be defused. The large moons in our Solar System (including our own) are also planemos, but not planets. Planetary scientists, as a practical matter, already study planemos rather than exclusively planets. I believe this scheme answers all the questions and requirements posed at the beginning, is relatively easy to implement, and could garner widespread support (if enough people are prepared to reconsider their traditional preconceptions). As scientists, I think we should remain consistent with our basically empirical and logical roots. I also believe that if we lead on this issue, the public will follow.

**Definitions**: *Planet: a planemo that orbits a fusor.*
*Planemo: a round non-fusor.*
*Fusor: an object capable of core fusion.*

One could of course drop the extra new words, and combine everything into a longer sentence. Legal disclaimers: "round" means "whose surface is very nearly on the gravitational equipotential"; "orbits" means "whose primary orbit is now, or was in the past around"; "capable" implies fusion is possible sometime during the existence of the object by itself.

## Closing Remarks

A number of scientists have offered opinions on the definition of "planet". Many, including the IAU Working Group on Extrasolar Planets, have accepted the fusion limit as setting the upper bound for planets. Some reject this because deuterium burning is weak at the boundary, or because they prefer cosmogony to an upper mass bound. The idea of something like planemos has also been suggested by Nick Wolff in the form of an acronym: PMOs (but with semi-arbitrary mass boundaries, and favoring cosmogony). Most reject free-floating objects as planets, but accept them as having "planetary mass" (although some stick with a purely mass-based definition and call them planets). The lower bound for planets is often suggested to be gravitationally-induced roundness, but arbitrary or other size limits that keep the Solar System census unchanged have some support. The IAU has tried to avoid this issue, but is being forced to revisit it by the discovery of 2003 UB313.

Many consider being a member of a belt population to be a disqualifying circumstance, and most of that group would drop Pluto's planetary status. Others would grandfather it in for tradition's sake, or refer to it with a dual status: smallest planet/largest KBO. Stern and Levinson (2002) have proposed something very similar to GB, but with a full range of astronomical adjectives transferring over: subdwarf, dwarf, subgiant, giant, supergiant. The subdwarf boundary comes where they calculate insufficient dynamical dominance, and their other boundaries are set at logarithmic intervals in Earth masses (corresponding roughly to terrestrial planets, ice giants, gas giants, and superplanets).

It appears in the end that well-defined boundaries, empirical verifiability, or logical consistency as requirements on a definition cannot necessarily overcome gut feelings (based on culture and tradition) that certain objects just don't belong in the category of "planets", while others just do. The bases of such feelings are not always the same, and they arise from the different weights given to the different arenas in which the debate is held. This is the essentially human nature of the problem. There do seem to be acceptable solutions, but agreeing on any one of them is difficult. We hope that this essay makes the problem clearer, and that substantial further reflection by everyone concerned, coupled with the ongoing flood of new discoveries, lights the path to a consensus. In the meantime, the debate itself keeps the field fresh and the public informed and engaged.